\documentclass[fleqn,usenatbib]{mnras}

\usepackage{newtxtext,newtxmath}

\usepackage[T1]{fontenc}
\usepackage{ae,aecompl}

\usepackage{graphicx}	
\usepackage{amsmath}	
\usepackage{amssymb}	
\usepackage{xcolor}
\usepackage[utf8]{inputenc}

\newcommand{\Msun}{\text{M}_{\odot}}	

\title[A deep regression network for halo formation]{Predicting dark matter halo formation in N-body simulations with deep regression networks}

\author[M. Bernardini et al.]{
M. Bernardini,$^{1}$\thanks{E-mail: mauro.bernardini@uzh.ch}
L. Mayer,$^{1}$\thanks{E-mail: lmayer@physik.uzh.ch}
D. Reed$^{1}$
R. Feldmann$^{1}$
\\
$^{1}$Center for Theoretical Astrophysics and Cosmology, Institute for Computational Science, University of Zurich,\\
Winterthurerstrasse 190, CH-8057 Zurich, Switzerland\\
}

\date{Accepted XXX. Received YYY; in original form ZZZ}

\pubyear{2019}

\begin{document}
\label{firstpage}
\pagerange{\pageref{firstpage}--\pageref{lastpage}}
\maketitle

\begin{abstract}
Dark matter haloes play a fundamental role in cosmological structure formation. 
The most common approach to model their assembly mechanisms is through N-body simulations.
In this work we present an innovative pathway to predict dark matter halo formation from the initial density field using a Deep Learning algorithm. We implement and train a Deep Convolutional Neural Network (DCNN) to solve the task of retrieving Lagrangian patches from which dark matter halos will condense. 
The volumetric multi-label classification task is turned into a regression problem by means of the euclidean distance transformation. The network is complemented by an adaptive version of the watershed algorithm to form the entire protohalo identification pipeline. 
We show that splitting the segmentation problem into two distinct sub-tasks allows for training smaller and faster networks, while the predictive power of the pipeline remains the same. 
The model is trained on synthetic data derived from a single full N-body simulation and achieves deviations of $\sim$10\% when reconstructing the dark matter halo mass function at $z=0$. 
This approach represents a promising framework for learning highly non-linear relations in the primordial density field. As a practical application, our method can be used to produce mock dark matter halo catalogues directly from the initial conditions of N-body simulations.

\end{abstract}




\section{Introduction}
The fundamental information source of gravitational dynamics is the cosmic density field described by its non-linear evolution. Cosmological N-body simulations of cold dark matter show how initially over-dense regions collapse through the competition of cosmic expansion and gravity to form virialized structures called dark matter haloes.
These haloes form the building blocks of large-scale structure as they define the landscape of potential wells in which baryonic matter flows to form galaxies, groups and clusters of galaxies \cite[e.g.][]{Wechsler2018, Guo2010}. 
The structure and formation of dark matter haloes is thus an important mechanism to understand when building a complete model of galaxy formation and evolution as intrinsic galaxy properties depend on the host dark matter halo mass and morphology \cite[][]{Wechsler2018, Feldmann_2019}.\\

The source of structure formation lies in the primordial perturbations seeded in the matter field of the early Universe, where initially small density peaks grow linearly through mass accretion and mergers of smaller structures. As the evolution of the matter density field progresses into the non-linear regime, the continuing gravitational accretion starts to form distinct virialized objects, that strongly affect their corresponding environments and substructures. As individual systems transition into the fully non-linear regime, the assembly history of dark matter haloes becomes generally difficult as complicated gravitational affects like merging or tidal stripping of larger structures can occur. In this regime, N-body simulations provide the only reliable tool for accurately describing structure formation processes.\par

The development of analytical approximations for describing where and how structure forms in an initial random field has opened the possibility to understand the main physical aspects that drive halo assembly.
The fundamental analytical work by \cite{Press_Schechter1974} suggests that dark matter collapse occurs once the spherically smoothed linear density field exceeds a cosmology-dependent threshold value. This idea is complemented by the excursion set formalism by \cite{Bond1991} which describes gravitational collapse as a statistical framework based on density trajectories on varying scales. This framework was further developed by relaxing the assumption of spherical collapse and calibrating with numerical simulations.
These semi-analytical schemes have been shown to reproduce halo statistics with acceptable error margins providing a fast method for sampling mock dark matter halo catalogues \cite[][]{Sheth2001, Reed2003}.\\

In recent time, a wide variety of machine learning techniques have been investigated to solve tasks related to non-linear structure formation \cite[e.g.][]{Berger2019, Calvo2019, He2018, LucieSmith2018, Zhang2019, Agarwal2018}. Generally, the approach of examining fully non-linear collapse dynamics by running N-body simulations is computationally expensive. Incorporating machine learning is therefore a natural step, since these algorithms are very adaptive to a large variety of problems and arrive at predictions comparatively fast.\par
Convolutional Neural Networks (CNNs) form a branch of Deep Learning models that have drastically changed computer vision problems \cite[][]{Krizhevsky2012}. As many other deep learning architectures, CNNs use the compositions of non-linear functions to model complex dependencies between input features and target variables. An interesting subsection of CNNs deals with the task of image segmentation, where networks are trained to identify and retrieve structures in input images. Through the combination of layered non-linear activation functions, CNNs learn derivatives in the input space to isolate objects of interest. Networks trained on segmentation tasks have a large variety of applications that range from identification of every day objects \cite[e.g.][]{Liu2015, Noh2015} up to nuclei segmentation in biological images of cancer tissue \cite[e.g.][]{Milletari2016, Naylor2019, Isensee2018}.\par
In this work we present an application of machine learning based on a CNN to identify Lagrangian patches (termed protohaloes) in the primordial density field, from which dark matter haloes condense. We formulate the problem in such a way that the pipeline can be used to produce mock dark matter halo catalogues directly from the initial conditions (IC) of N-body simulations. We design the classification task of distinguishing between collapsing and background regions as a regression problem in the distance space. This alternative technique was first proposed by \cite{Naylor2019} and shows great success for images with overlapping regions. The precision of the entire pipeline is measured by comparing the reconstructed dark matter halo mass function to its corresponding ground truth.\par
Since identifying collapsing regions in the primordial density field is a highly non-trivial task, the success of training a machine learning model strongly depends on how this mapping is formulated numerically.
\cite{LucieSmith2018} trained a Random Forest on data retrieved on a particle by particle basis to predict whether or not a particle will be a member of a dark matter halo at $z=0$. They successfully showed that the primordial density field contains enough information to predict halo membership as the precision of their algorithm outperformed analytical frameworks like the Extended Press-Schechter formalism. They also showed that the density contrast is an information carrier to predict the final distance of particles from their corresponding halo density peaks.\par
\cite{Berger2019} presented a powerful binary classification network (that inspired our own implementation) and paired it with an algorithm to extract halo regions from the reconstructed probability map. Compared to the particle information in \cite{LucieSmith2018}, their neural network operates upon gridded density contrast information. Regarding halo formation, this approach has the advantage that the network learns to identify the important features in the input map and derivatives thereof itself, rather than being presented with a fixed feature vector as in \cite{LucieSmith2018}. This allows the network to scan the exact shape of the local neighborhood around density peaks on different scales, in order to assess the underlying collapse dynamics. The downside of formulating halo membership in a binary approach is, that retrieving individual haloes strongly depends on border pixels having smaller probability values then centrally located cells, in order to minimize the undesired effect of haloes artificially clumping together. This indicates that the success of this method relies on the imperfection of the neural network prediction regarding the reconstructed probability map as halo borders are identified by varying probability thresholds. \par
Inspired by this approach, we seek to formulate a mapping where the network is trained on a target that incorporates the information of protohalo borders by construction. 
We deliberately focus on retrieving halo information from the initial conditions rather than from the $z=0$ snapshot for two very important reasons. First, the central location of haloes changes over time through gravitational interactions as they gradually divert from the initial locations of the primordial density peaks.
Second, the final halo sizes at $z=0$ in full N-body simulations are small compared to the box size and thus span only a very small subsection of the simulation volume, making the target field sparsely populated. Training on sparsely populated target fields of the exact halo position and mass is a very difficult task as has been pointed out by \cite{Zhang2019} (although in a slightly different application). For these reasons we choose to design the network mapping where input and target both display information at the initial conditions.
We show that by introducing gradient fields in the target map, and in particular the distance information, one can train the network to predict membership of central pixels more accurately than border cells. This gives us the advantage of well-defined borders in the target map, as information of individual objects is retained.\par
This paper is structured as follows. In section \ref{sec:DL_structure} we discuss the methods we used to construct the training samples for the supervised regression problem from cosmological N-body simulations. Then, we outline the mapping the network is trained on and in section \ref{sec:NN_implementation} we present our network implementation and training strategy in great detail. In section \ref{sec:dist_to_seg} we describe an adaptive algorithm to reconstruct the segmentation map from the network output, in order to predict the dark matter halo distribution for a single simulation box. We assess the network precision by comparing the predicted and true halo statistics in form of the halo mass function at $z=0$. We furthermore discuss distinct strengths and weaknesses of the network at its current state when confronted with specific cases of density configurations. We propose future ideas for further improvements and conclude in section \ref{sec:conclusion}.

\begin{figure*}
 \includegraphics[width=0.75\textwidth]{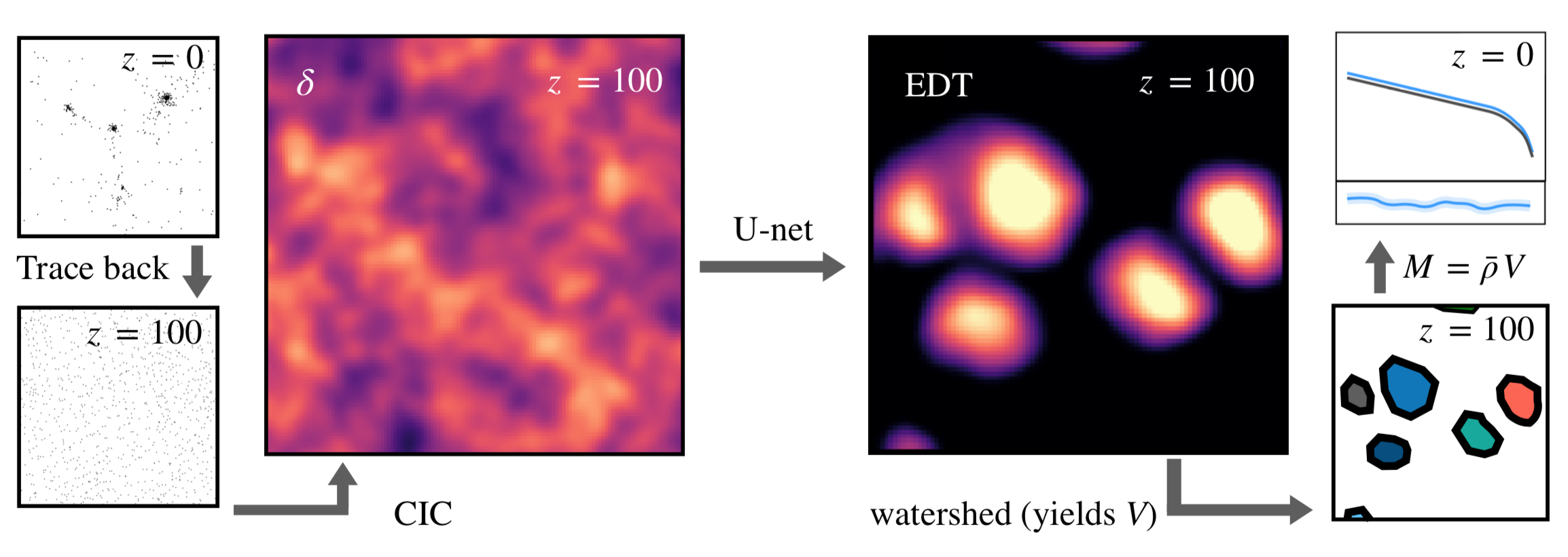}
 \caption{A schematic overview of the entire pipeline. We describe the network mapping and how the training samples of density contrast $\delta$ and EDT (defined in eq. \ref{eq:distance}) are constructed in section \ref{sec:DL_structure}. We discuss the network architecture, training strategy and results in section \ref{sec:NN_implementation}. In section \ref{sec:dist_to_seg} we introduce the watershed algorithm as a key element in the post-processing of the neural network predictions and describe how the halo mass function is reconstructed.}
 \label{fig:pipeline}
\end{figure*}
\section{Deep Learning for structure identification}
\label{sec:DL_structure}
In this section, we present the details of our implemented pipeline in the following sequence:
\begin{itemize}
    \item We present the numerical details of the N-body simulations used in this work as well as the identification algorithm of dark matter haloes at $z=0$.
    \item We introduce the concept of a protohalo at the initial conditions, how these objects are retrieved from the simulation and how they are used in the prediction algorithm.
    \item We describe the numerical concept of the distance transformation as a metric in the context of protohalo boundaries and how this information allows to construct the data samples for the network mapping.
\end{itemize}

\subsection{Cosmological N-body simulations}
We adopt two N-body simulations of the same $\Lambda$CDM cosmology ($\Omega_{m}=0.279, \Omega_{\Lambda}=0.721$) produced by \texttt{pkdgrav} \citep{pkdgrav2000}, one for training and one for testing the algorithm. The comoving box size is 100$\,h^{-1}$Mpc with a particle resolution of $512^{3}$, where each individual dark matter particle carries a physical mass of $8.84 \times 10^{8}\,\Msun$.
We use the \texttt{HOP} halo finder by \cite{Eisenstein1998} to retrieve dark matter halo catalogues from the simulations. This group finding algorithms first associates a density estimate for all particles. Each particle is then iteratively linked to its densest nearest neighbor until the algorithm reaches a particle which is its own densest neighbor. The collection of particles that are traced to the same densest particle forms a single group.
In this way, there are no constraints on the morphology of individual haloes and since this method does not rely on a linking length, the undesired effect of bridging discrete haloes together is avoided entirely. We note that the neural network mapping is unaltered when choosing a different halo finder as the network learns the concept of a protohalo from the data samples itself.\par
We chose a lower halo mass threshold of $4\times 10^{12}\Msun$ (corresponding to 4525 particles) as smaller haloes are not sufficiently resolved for the underlying task. Running the halo finder over the corresponding $z=0$ snapshots results in catalogues of 6D phase-space information for each halo. The training simulation contains 1418 dark matter haloes ranging up to $\sim$10$^{15}\Msun$ where the validation simulation contains 1630 thereof.

\subsection{Formulating the network mapping}\label{sec:network_mapping}
The primordial density field is the source of input information from which the network learns to retrieve the collapsing protohalo regions. 
We define a protohalo as the collection of particles in the initial conditions that will all end up in the same halo at $z=0$. The entire particle set at the ICs can therefore be split into background particles (that will not be part of any virialized structure) and multiple progenitor clumps that will collapse into distinct dark matter haloes.\par
Due to the fact that CNNs operate upon gridded data, we switch from the spatial particle distribution to a grid-based density contrast by means of the cloud-in-cell algorithm \cite[e.g.][]{Howlett2015}. The grid resolution is chosen to be $512^{3}$, so that each cell contains approximately one particle at the initial conditions. We transform the deposited density contrast
\begin{equation}
    \delta = \rho / \bar{\rho} - 1
\end{equation}
by unit-variance scaling, where the data is divided by the standard deviation of all pixel values. The distribution of the resulting dataset is centered around 0 with a standard deviation of 1. This data pre-processing step has been shown to help in faster convergence of deep learning models \citep{LeCun1998}. \par
\begin{figure}
    \begin{minipage}[t]{0.5\columnwidth}
        \vspace{0pt}
        \includegraphics[width=\columnwidth]{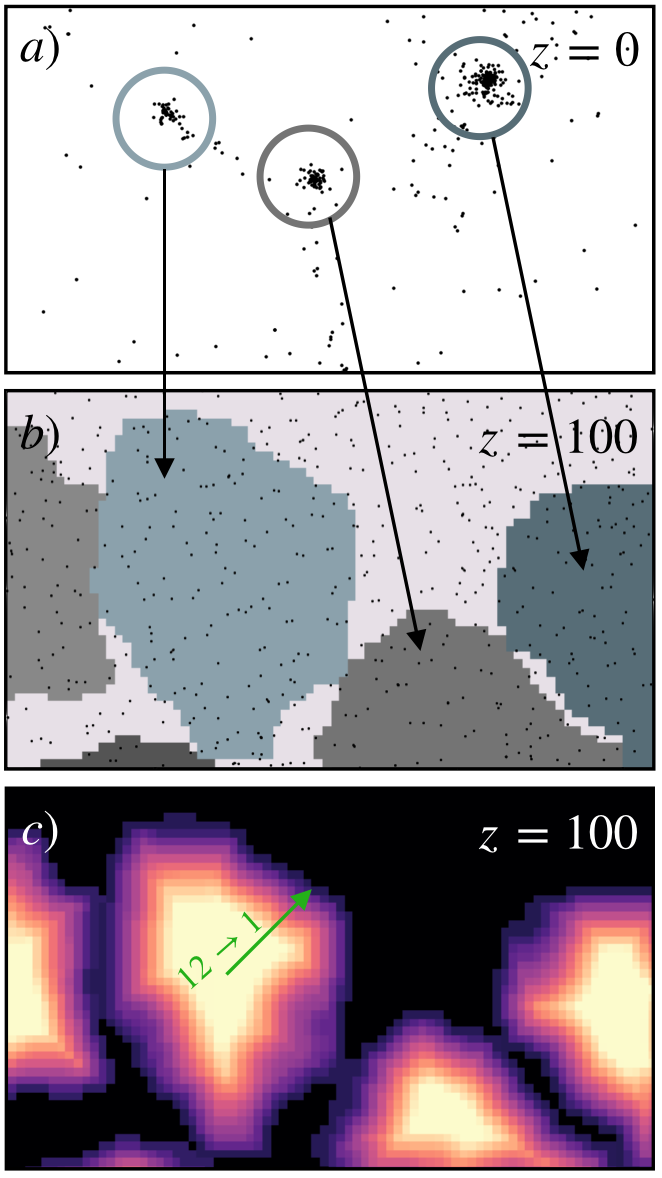}
    \end{minipage}\hfill
    \begin{minipage}[t]{0.47\columnwidth}
        \vspace{0pt}
        \caption{An example situation of the target sample generation pipeline. Particles in haloes at $z=0$ are traced back to the initial conditions where the associated halo mass is deposited hierarchically as described in section \ref{sec:network_mapping}. This results in distinct protohalo regions marked by different colors to better convey individual structures. We then apply the EDT transformation to the entire label image and normalize on an object by object basis to produce the distance map $B_{d,n}$ shown in the third subplot. Also shown in green is the actual and normalized distance value (12 and 1) to the closest background cell for the innermost cell for an example protohalo. Background cells (shown in black) have zero distance. The normalized euclidean distance information conveys that the target map retains individual objects even though distinct patches might be connected.}
        \label{fig:overview}
    \end{minipage}
\end{figure}

\cite{Naylor2019} presented a novel approach to turn a pixel-wise classification problem into a distance-based regression task.
Let us assume that an image harbors multiple well-defined objects defined by different labels as seen in figure \ref{fig:overview} (subplot $b$). All pixels are associated a label that is unique of the enclosing object they belong to, whereas pixels not belonging to any substructure typically carry a zero label. We define the set $K_{x}$ as the collection of pixels $x$ that belong to a specific object in the image. The residual cells denoted as $y$ define the background of this substructure, i.e. $y\not\in K_{x}$. The euclidean distance transformation (EDT) of a multi-labelled map $B$ is then defined as $B_{d}=d(B)$, where each cell $x$ is assigned the euclidean distance to the closest background pixel as defined above, i.e.
\begin{equation}
    d(x)=\text{min}_{y\not\in K_{x}}{\left| x-y \right|}.
    \label{eq:distance}
\end{equation}
The transformation is conducted with the \texttt{edt}\footnote{\href{https://github.com/seung-lab/euclidean-distance-transform-3d}{\texttt{github.com/seung-lab/euclidean-distance-transform-3d}}} package, which implements euclidean distance transformation for multi-label regions.
The main problem in binary segmentation is that close or overlapping objects tend to be segmented as one single region.
Instead of predicting the protohalo membership of individual pixels in a binary fashion, each cell is assigned the nearest distance to the border of the cluster it resides within.\par
We therefore construct the distance map of protohalo regions in the following way. Apart from general 6D phase-space information, each dark matter halo in the catalogue also carries the unique ID's of its member particles. Conversely, we record for each particle the mass of the halo it resides in at $z=0$. If a particle does not belong to a halo it is assigned a halo mass label of 0.
This mass information is then traced back for each particle to the initial conditions at $z=100$, where it is deposited in the following hierarchical fashion. Each cell of the target map is assigned the largest halo mass label present inside it. If multiple labels are present in a single cell, the deposited value corresponds to the mass of the largest halo, where cells with no halo particles inside are assigned a zero-value corresponding to the background. This top-down approach prioritizes large scale protohaloes by construction, but since at $z=100$ approximately one particle resides in each cell, almost all of the halo information is preserved. Moreover, as the labels are directly retrieved from the halo catalogue at $z=0$, the masses of dark matter haloes and their corresponding protohaloes equal in very good approximation.
The entire pipeline is schematically shown in figure \ref{fig:overview}.\par
Regressing the distance information as the mapping target offers the major advantage that the networks attention is primarily drawn to the central cluster regions where the density contrast is generally largest. As outlined in section \ref{sec:training}, the training loss function compares the predicted and target distance values in a direct way, such that the prediction on central cells is more weighted compared to border pixels. Additionally, as the dark matter haloes mass ranges from $4\cdot 10^{12}\Msun$ to $\sim$10$^{15}\Msun$ the size of regions, and thus the distance values, vary across different scales. We therefore decide to normalize the distance map of each individual protohalo cluster, such that the value of the central cell equals 1 for all objects. The individually normalized distance map $B_{d,n}$ forms the regression target in the network mapping. We also investigated training on the target of non-normalized distance maps and found that the network primarily focuses on predicting the correct distance value for large scale objects while often failing on small scales. We show a sample result of these operations in figure \ref{fig:overview}.

\subsection{Generating synthetic training samples}
We construct the training and validation samples from the two deposited particle fields described above. The input $i$ is the unit-variance scaled density contrast whereas the ground target $g$ corresponds to the normalized distance map. The network is trained on regressing the correct distance value from the density contrast, where both fields represent the state at the initial conditions. The data pairs $(i,g)$ are constructed by a tiling strategy, which is similar to the one presented by \cite{Berger2019}. We divide the entire $512^{3}$ domain into subvolumes of $128^{3}$ where adjacent samples overlap by 64 cells to avoid edge effects. In this manner, the two simulations are each divided into 512 samples, where the innermost $64^{3}$ cells are unique to each subvolume. 
Each cube can be rotated in 6 random directions (2 per axis), where for each direction exist an additional 4 possibilities to orient the cube around the given axis, giving a total augmentation factor of 24. We augment the training and validation sets by the aforementioned transformations resulting in a total set of 12'288 samples for training and validation respectively.

\section{Neural network implementation}
\label{sec:NN_implementation}
CNNs achieve feature extraction from samples by convolving the inputs with filters to produce so called feature maps. The information stored in these feature maps is subsequently down-sampled by pooling operations, which are designed to preserve the important information signals. As data descends deeper into the network, the feature representations generally grow in numbers, while the size of the feature maps decreases due to information pooling. The network architecture is designed in a way that the information of individual feature maps flows together inside deeper convolution layers, allowing the network to learn non-linear combinations of identified features.

\subsection{U-net architecture}
\begin{figure*}
 \includegraphics[width=\textwidth]{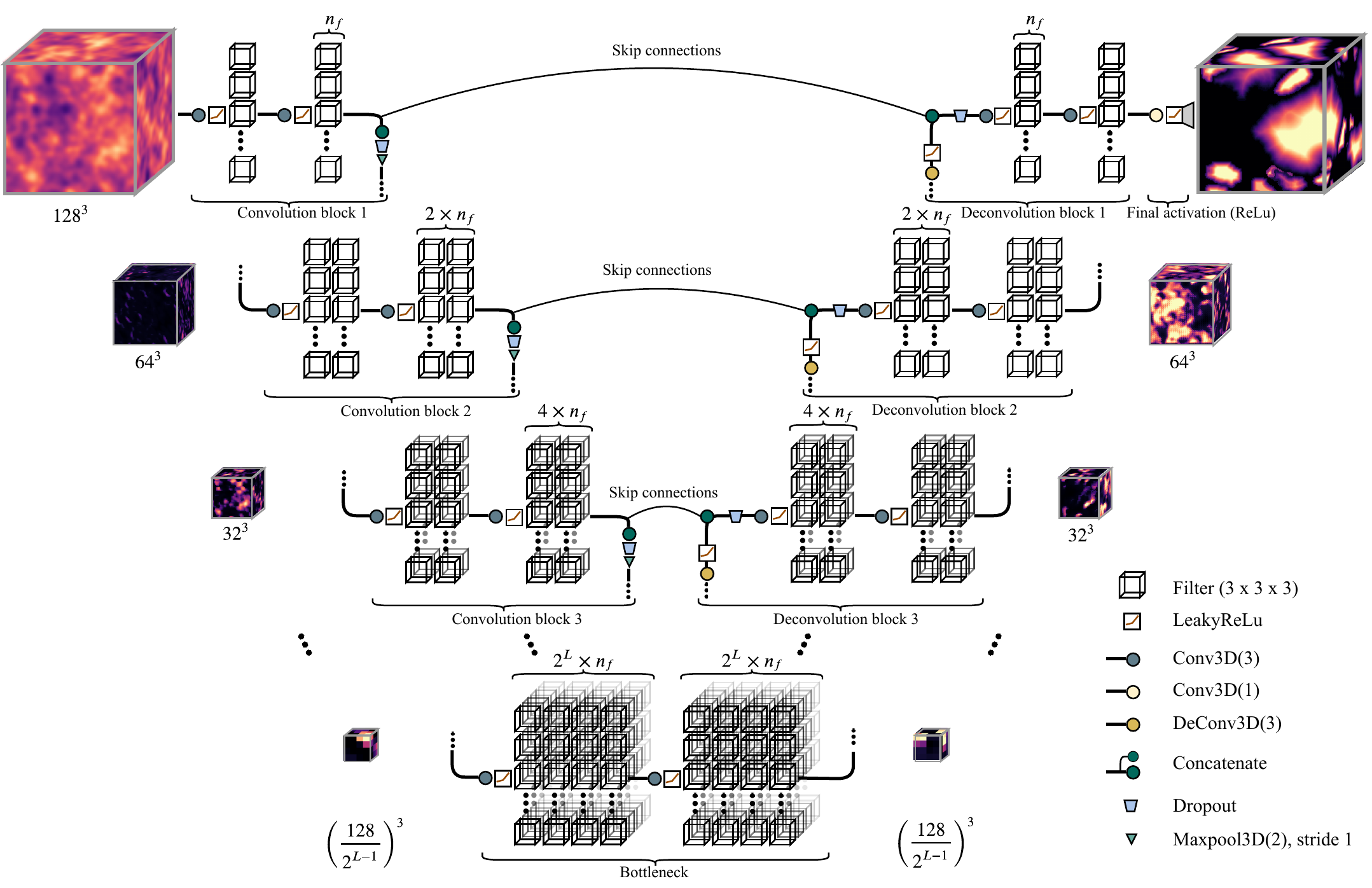}
 \caption{A schematic representation of the fully convolutional U-net implementation used for the regression task. The network is constructed by stacking and connecting convolution blocks down to a bottleneck layer, from which the original input dimensions are reconstructed by upsampling the data through deconvolution blocks. The skip connections concatenate the convolution and deconvolution blocks with same dimension in every corresponding network layer. Also shown are the intermediate layer outputs before each convolution block and after each deconvolution block respectively. With each downward step, the dimensions decrease by a factor of 2 due to the MaxPooling operations, whereas in the decoder part the transpose convolutions upsample the dimension again by the same factor of 2.}
 \label{fig:network}
\end{figure*}

In this work we make use of the U-net first introduced by \cite{Ronneberger2015} for solving bio-medical image segmentation tasks. The network itself is a fully-convolutional autoencoder consisting of two main branches, an encoding and decoding part. As in the general autoencoder case, the information compression is realized by convolutions followed by pooling operations. For image segmentation the output and input dimensions must match. In order to recover the original input dimensions, the decoding blocks are constructed by a single transpose convolution followed by consecutive convolution operations, which results in upsampled feature maps. A schematic depiction of the entire network is shown in figure \ref{fig:network}, where the data flows in a U-shape form from the upper left to the lower middle and then reascends along the right hand side. In this process the network is able to extract advanced features, but looses localization information at the same time. In order to correct for this loss, \cite{Ronneberger2015} introduced skip connections (often termed \textit{fine-grained feature forwarding} connections), that copy and concatenate the information from the corresponding encoder level with the up-flowing data in the decoder part. In this manner, the spatial information from the contraction path is directly transferred to the expanding branch without being passed through the bottleneck and deconvolution operation. Additionally, \cite{Ronneberger2015} also found that the skip connections greatly reduce training time regarding model convergence.\par
We implement and train our own U-net version in \texttt{keras}, a high-level neural networks API \citep{Chollet2015}.
The individual convolution blocks are constructed by two subsequent convolutions of filter size $(3\times 3\times 3)$ and stride 1. The first network layer deploys $n_{f}$ convolution filters, where we choose $n_{f}$=12 in our implementation. After each convolution a non-linear activation function is applied to the data tensors. We choose LReLu, a leaky variation of ReLu (Rectifier-Linear-Unit), with an $\alpha$-value of 0.05. The main reason for this choice is that deep networks with native ReLu activations have been found to occasionally suffer from vanishing gradients \cite[e.g.][]{bengio1994}. The main advantage of LReLu lies in its non-zero gradient, which makes the network more robust as training should in principle never stop.
After the second activation in each layer the data is copied and split along two paths. The first path is the skip connection, which concatenates to the up-flowing data on the reascending network part. Along the second path the data flows through a Dropout layer \citep{Srivastava2014} and finally passes through a single MaxPool operation with filter size $(2\times 2\times 2)$ and stride 1. As the data descends the network reaching deeper levels, the architecture of the convolution blocks remains the same with the exception that a total of $(2l\times n_{f})$ filters are applied, where $l=[1,...,L]$ is the corresponding layer number. The final network is constructed with a total of 5 convolution blocks (i.e. $L=5$). For deeper layers the network becomes more sensitive to large scale structures in the input image since the pooling operations cut the corresponding image dimensions in half for each additional layer. As the physical size of one simulation cell is $\sim$0.2 Mpc, the network downsampling increases the receptive field by a factor of $2^{5}=32$. The deepest network filters of size $(3\times 3\times 3)$ can therefore learn $\sim$20\,Mpc features in the density field. In the upsampling branch features of all scales are used in conjunction with the spatial information provided by the skip connections to assemble the final prediction. We choose a native ReLu for the final activation function as the target only contains values $\geq\,$0.
\begin{table}
	\begin{tabular}{llr}
    U-net ($L=5, n_{f}=12$)                 &                          & \multicolumn{1}{l}{} \\ \hline
    architecture/operation & \multicolumn{2}{c}{occurrence}                  \\ \hline
                           & no. per layer $l$           & total no.            \\
    \texttt{Conv3D}                 & $2^{l+1}n_{f}$                   & 3024                    \\
    \texttt{Conv3DT}                & $2^{l}n_{f}$                       & 372                    \\
    \texttt{MaxPool}                & 1 (encoder)              & 5                    \\
    \texttt{Dropout}                & 1 (both)                 & 10                    \\
    \texttt{LReLu}                  & 2 (encoder), 3 (decoder) & 27                    \\
                           &                          & \multicolumn{1}{l}{} \\
    Total free parameters  & $14.5 \cdot 10^{6}$                        & \multicolumn{1}{l}{} \\ \hline
    \end{tabular}
	\caption{Summary table of all network hyper-parameters and layers for a 5-level version of the U-net with $n_{f}=12$ initial filters and filter size of $(3\times 3\times 3)$ throughout the network. Only \texttt{Conv3D} and \texttt{Conv3DT} contain trainable parameters.}
\end{table}

\subsection{Training the neural network}
\label{sec:training}
The training is conducted on individual batches of size 16 (train-on-batch strategy). For loss function minimization we use the Adaptive momentum optimizer (Adam) provided in the \texttt{keras} \citep{Chollet2015} package. By the gradient descent algorithm, the network learns relevant relations to accurately predict the distance map from the input density field. However, since a training sample only covers an area of $128^{3}$ cells, one expects the network to be less accurate when predicting the target value for pixels close to the outer border, because important parts of the density field are stored in the next adjacent subbox. Hence, the approach of splitting the entire simulation domain into subboxes introduces an obstacle in the training algorithm caused by edge effects. To correct for this, we implement a custom loss function based on the normalized $L_{1}$ loss (normalized mean-absolute-error), that is only sensitive to pixels inside a predefined receptive field. The buffer size is chosen to be 16, so that only the innermost $96^{3}$ cube is considered when evaluating the cost function. We choose a smaller buffer size compared to the buffer of the individual subboxes, because otherwise larger structures would not entirely fit inside the receptive field of the loss function and their identification would have to span across neighboring subboxes. This is undesired, since the neural network is only presented with a single instance during prediction time. Formally, the selective loss function is constructed as
\begin{equation}
    \hat{L}_{1,\text{sel}} = \frac{1}{n\bar{g}} \sum_{k}^{n}(g_{k}-p_{k})^{2} \Big|_{\Box 96^{3}} 
\end{equation}
where $g_{k}$ and $p_{k}$ denote individual cells of ground truth and prediction. We choose to normalize the loss function by dividing by the mean ground truth distance $\bar{g}$, since different subboxes show largely varying distance values depending on how many collapsing regions are located inside it. The dropout rate is set to 0.5 during the entire training period.\\

\begin{figure}
 \includegraphics[width=\columnwidth]{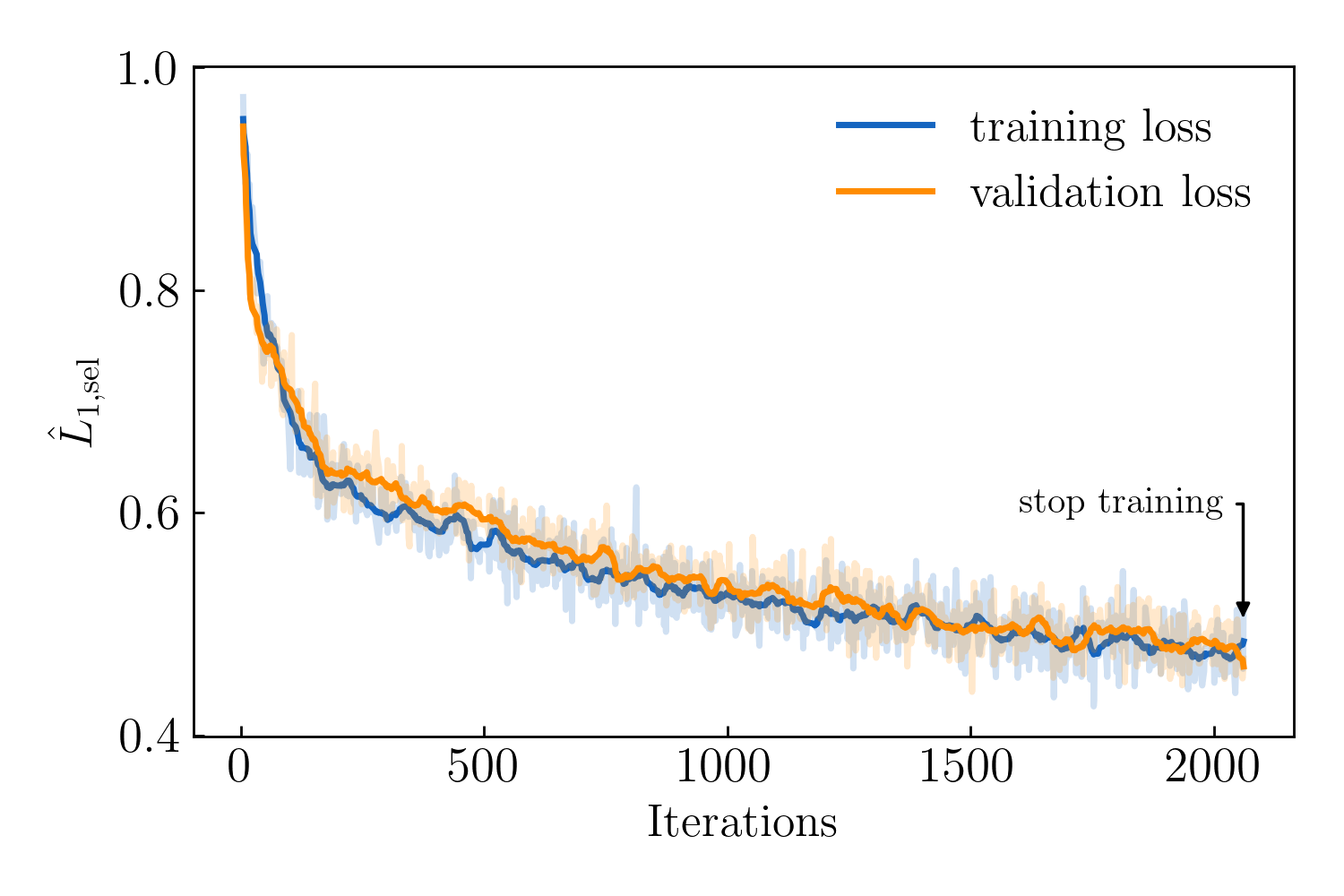}
 \caption{The evolution of the smoothed loss function $\hat{L}_{1,\text{sel}}$ for both training and validation simulation over the entire $\sim$2000 training iterations, after which the loss improvements become marginally small and training is stopped. The dropout rate is set to 0.5 throughout the entire training process.}
 \label{fig:loss_evolution}
\end{figure}

In figure \ref{fig:loss_evolution} we show the evolution of the loss function, where after $\sim$2000 iterations loss minimization stagnates and training is stopped. Despite the comparatively low number of training samples we see that the current strategy is successful as the loss steadily decreases with no overfitting occurring.
The network reaches a loss value of $\sim$0.6 in a relatively short training period. The subsequent gradient descent however is slow resulting in the characteristic exponential tail that is often encountered when training deep learning algorithms. The small training set of only 512 unique samples presents another challenge as the global minimum may be surrounded by local minima that are characteristic for the training set. Moreover, the comparatively small batch-size of 16 as well as the train-on-batch strategy make the gradient descent less smooth. Even though in principle this can be overcome by changing the training strategy, the shortcoming is the loss of quick access control to the training algorithm because the network weights do not get updated as frequently as in the train-on-batch strategy.\\

\begin{figure*}
 \includegraphics[width=\textwidth]{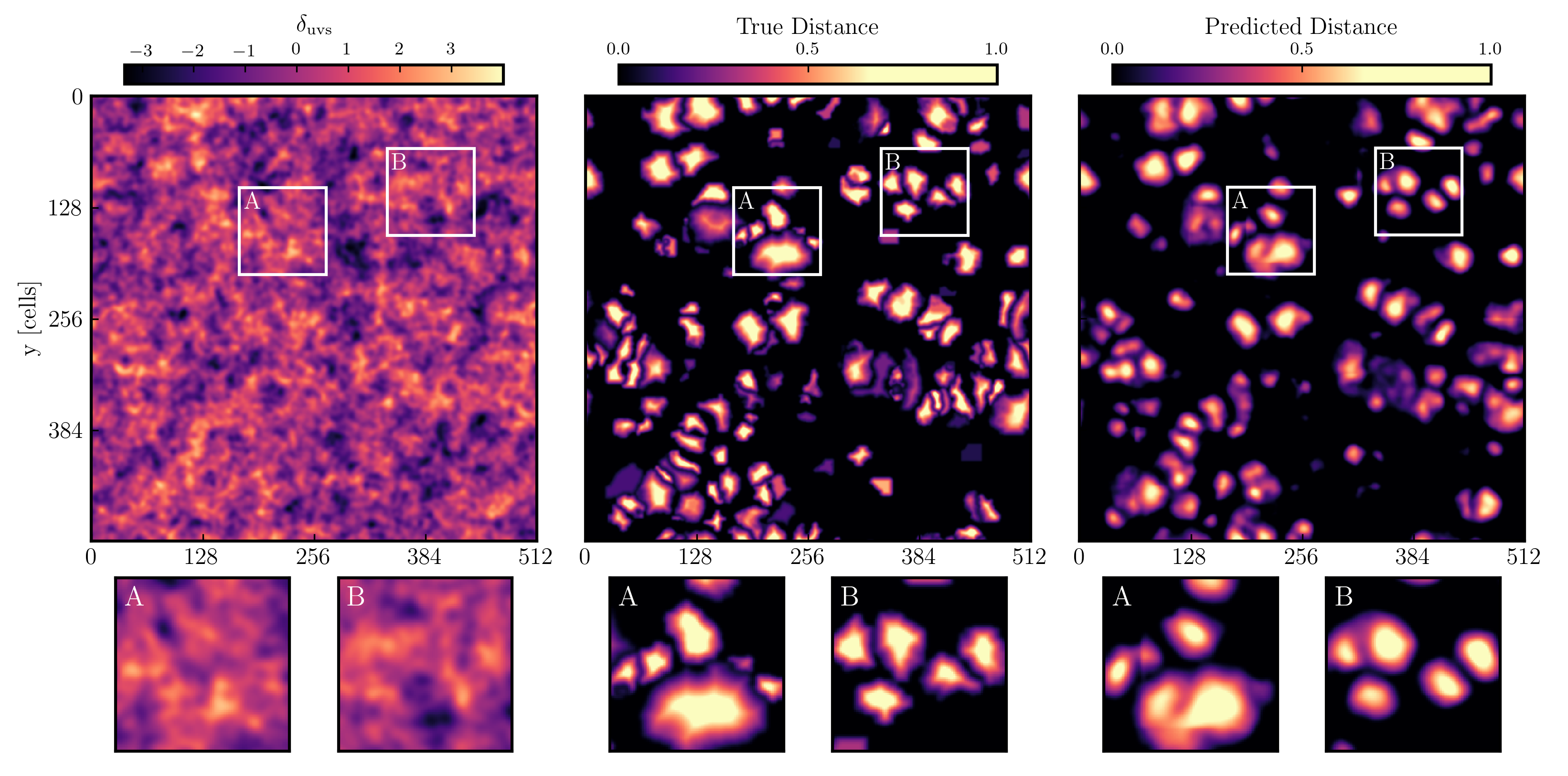}
 \caption{An entire slice of the validation box displaying the input density contrast at $z=100$ (left), raw and unprocessed prediction of the neural network (right) and ground truth (middle) of the normalized distance map. 
 Also shown are insets of two regions conveying the performance of the neural network in more detail.
 In case B the network manages to predict the exact number of distinct clusters as well as a good estimate on the corresponding sizes, whereas case A shows a problematic prediction regarding protohalo identification where a major density patch is split into two sub-structures. This behaviour is unwanted as the reconstructed region will suffer from oversegmentation.
 }
 \label{fig:triple_combined}
\end{figure*}

The trained network is then used to predict the distance map for the validation simulation. A slice of the validation box is displayed in figure \ref{fig:triple_combined}. In general, the network manages to identify the important regions of gravitational collapse and to estimate the corresponding protohalo sizes. For large and intermediate scales the general sizes of connected structures in the distance map agree well with the ground truth. There are however individual cases where the compact and intermingled structure of many nearby clusters poses a great challenge for the network as it is not able to accurately split the individual regions in the correct way. Occasionally, two intermediately sized protohalo regions will be merged by the network to produce a single cluster and vice versa. We show an example of such a faulty network prediction in the zoomed-in subplot A in figure \ref{fig:triple_combined}, where the network does infact predict the right size of the collapsing regions but splits the density patch into two sub protohaloes. Accurately predicting the distances of small scale structure poses the largest challenge for the network as there are at least three possible interplaying reasons for this behavior:
\begin{enumerate}
    \item The lack of numerical resolution induces signal-to-noise ratios in the density field that are too low for the network to retrieve any collapse information on small scales.
    \item The corresponding region in the input density field does in fact show signs of an overdensity but through gravitational dynamics the small cluster is either dispersed or merges with a larger nearby cluster and its isolated signal is lost.
    \item Regions that will form halos at $z=0$ with masses slightly below the halo mass barrier of $4\cdot10^{12}\Msun$ will not appear in the ground truth distance map, even though the signal is present in the density field at the initial conditions. This is a natural shortcoming of imposing a hard threshold on small scale structure.
\end{enumerate}

\section{From distance information to segmentation maps}
\label{sec:dist_to_seg}
In the following section we describe how the output from the neural network is post-processed to predict the mass of individual protohalo regions and to construct the final halo mass function.\par
Having trained the network to predict the desired distance information from the density contrast, we construct an adaptive algorithm that retrieves the boundaries of individual protohalo regions as defined by the metric in equation \ref{eq:distance}.
We make use of the watershed algorithm, a fast and reliable tool used in digital image processing for segmenting data with overlapping regions. The name refers metaphorically to a geological watershed, which separates adjacent drainage basins, as it treats the image it operates upon like a topographic map. With the brightness of each pixel representing its height (or distance as defined in equation \ref{eq:distance}), the algorithm finds the lines that run along the tops of ridges and is able to separate adjacent regions from one another. We suggest that the interested reader consult \cite{Kornilov2018} for a detailed overview. \par
We adopt the marker-based version of the watershed algorithm, which is extremely useful for this purpose as the distance map is the only needed ingredient.
In a preliminary step, the algorithm runs a local-maxima-finder to identify the positions of distance peaks, which play the important role of preselected markers. In a subsequent step, sources are placed at the marker positions, from which the image is flooded until water basins attributed to different markers meet on watershed lines. The resulting set of segmented regions constitutes a watershed segmentation by flooding.\\

\subsection{Generating halo catalogs and bias correction}
\begin{figure}
 \includegraphics[width=\columnwidth]{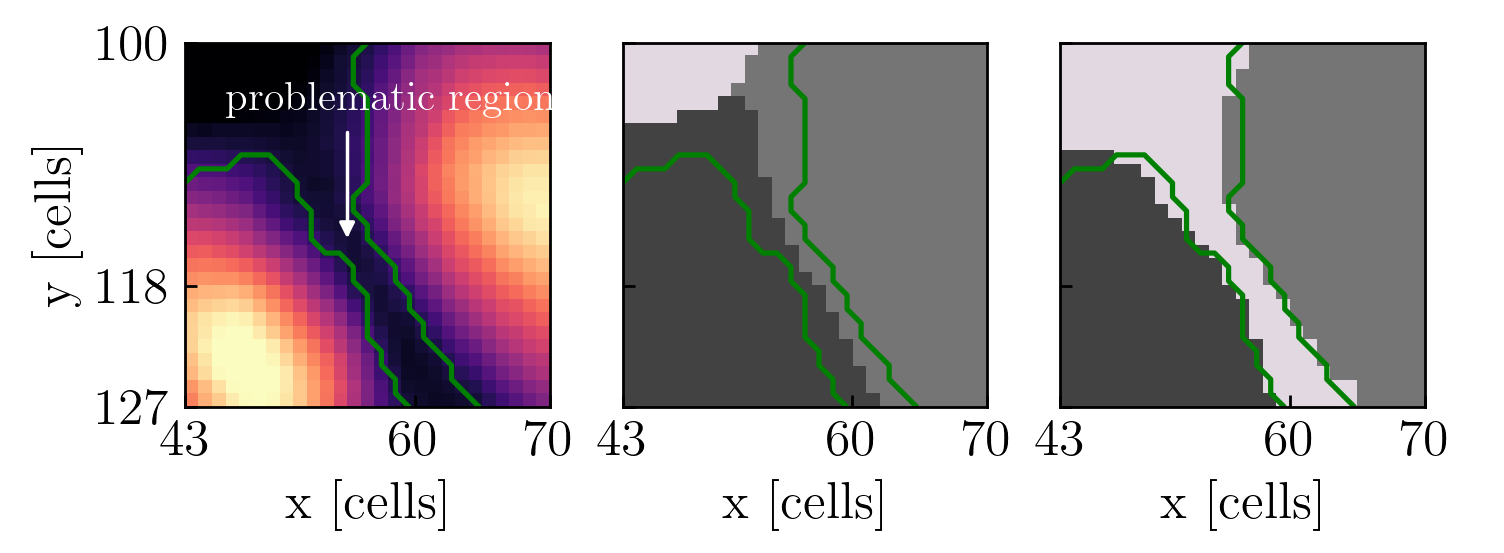}
 \caption{An example protohalo region that conveys very well the need for an adaptive version of the watershed algorithm. On the left a zoomed in slice of two predicted clusters is shown. The middle pane shows the raw cluster reconstruction from the native watershed algorithm, whereas the result of the adaptive post-processed version is shown on the right. Given the true protohalo boundaries overlayed in green in all three plots, it is evident that the adaptive watershed algorithm yields better estimates regarding the individual sizes of neighboring clusters, as problematic regions are filtered away.}
 \label{fig:adaptive_example}
\end{figure}

The native watershed algorithm however has a natural shortcoming when reconstructing images of structures with different scales. An example is shown in figure \ref{fig:adaptive_example}, where two clusters of different sizes are very close to one another. The middle pane shows the reconstruction from the native watershed algorithm. The general problem is that by simply flooding the image, any region where pixels have nonzero values is filled and gets assigned to one of the two clusters, which generally results in overestimated sizes when compared to the ground truth protohalos. This behaviour introduces a bias that changes with mass scale as it originates from the imperfect network predictions at cluster borders.\par
\begin{figure}
 \includegraphics[width=\columnwidth]{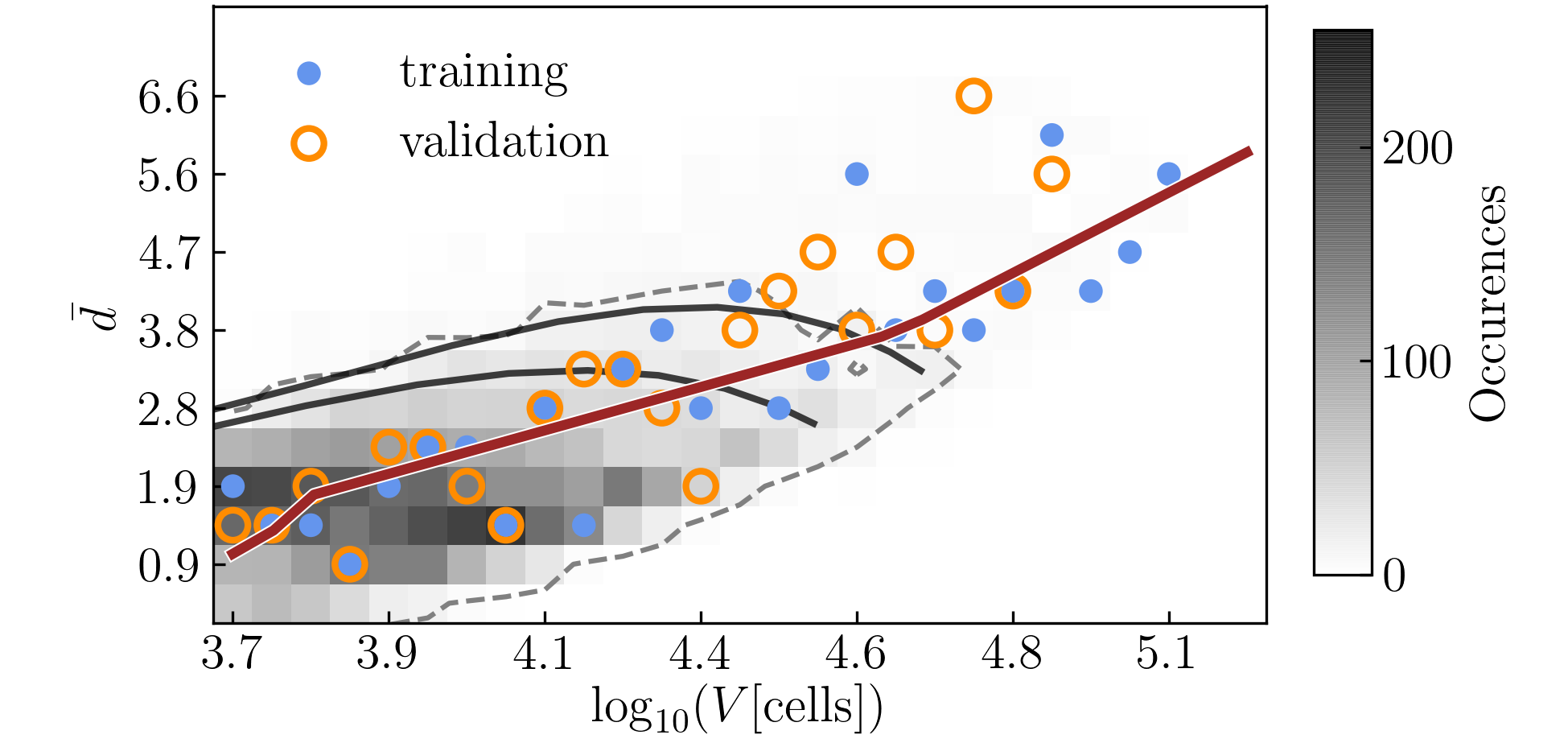}
 \caption{We show two example trajectories in the ($V,\bar{d}$)-space of two individual haloes in black, where increasing the contour threshold results in smaller and thus less massive regions.
 Beyond the dashed contour individual bins contain less than 10 intersections, implying very low statistics. Calibrating with the ground truth halo mass function yields the scatter markers where blue dots and orange circles represent training and validation data, where it is apparent that the overestimation follows the same trend in both datasets. We manually fit a piece-wise linear function (red) through the training scatter to retrieve the adaptive relation between $V$ and $\bar{d}$. The corrected protohalo sizes then correspond to the bin where their trajectories intersect the fitted relation.}
 \label{fig:fit}
\end{figure}

We correct for this overestimation by introducing an adaptive element to the native watershed output that restricts the regions, within which the algorithm can fill the image. This is achieved by thresholding the distance map of each individual cluster with an adaptive value that is found by calibration with the ground truth halo mass function in the following strategy. We switch back to the unnormalized cell-based distance $d$ by means of a spherical approximation for each cluster,
\begin{equation}
    d = \hat{d}\cdot\frac{3}{4\pi}V^{1/3},
\end{equation}
where $V$ denotes the uncorrected size in terms of individual cells and $\hat{d}$ is the normalized distance value from the prediction. By thresholding the distance map of a cluster with increasing values, we trace the evolution of the mass and the averaged distance values inside it. This results in unique trajectories in the ($V,\bar{d}$)-space for every protohalo region (see figure \ref{fig:fit}). 
A rapid change in mass, while the mean distance remains constant is an indication for a smeared border that is not well defined. The information of all trajectories is accumulated in a 2-dimensional histogram that represents the occurrences of intersections in each ($V,\bar{d}$)-segment. 
The ground truth halo mass function yields the information of how many halos are situated in a given size bin.
Through direct calibration we then identify for each mass scale, the distance bin that offers the smallest residual in terms of halo number with respect to the ground truth. The identified bins (shown as blue data points in figure \ref{fig:fit}) are then fitted with a piece-wise linear function to understand the bias introduced by watershed as a corrective relation between $V$ and $\bar{d}$. We are constricted to choose a comparatively low number of bins, as the small sample number of $\sim$1500 haloes does not offer enough statistics and is unreliable for finer binning.\\ \\
We implement the entire adaptive element as a post-processing step following the native watershed flooding, meaning that the entire cluster size retrieval includes the following two steps. First, the segmented image is reconstructed by the native watershed algorithm where sizes are generally overestimated. In a second step, the different reconstructed regions are thresholded depending on the mass bin where their corresponding trajectories intersect the piece-wise linear fit.
The numerical reason behind this approach is that the correction remains minimal because the actual masses are held as large as possible since only the outermost borders are cut away.
In this manner we construct an algorithm that is sensitive to different cluster scales as small and intermediate scale structures are prevented from being merged with nearby clusters. We find that this adaptive thresholding algorithm recovers the individual sizes much better compared to a uniform threshold value, which would suffice in the case of the neural network achieving close to perfect precision.
An example of this correction is showed in figure \ref{fig:adaptive_example}.
We emphasize that this step is purely due to the imperfections of the network predictions. Training deeper models with larger training sets at higher resolution may eradicate the need of this adaptive post-processing element entirely.\par
\begin{figure*}
 \includegraphics[width=\textwidth]{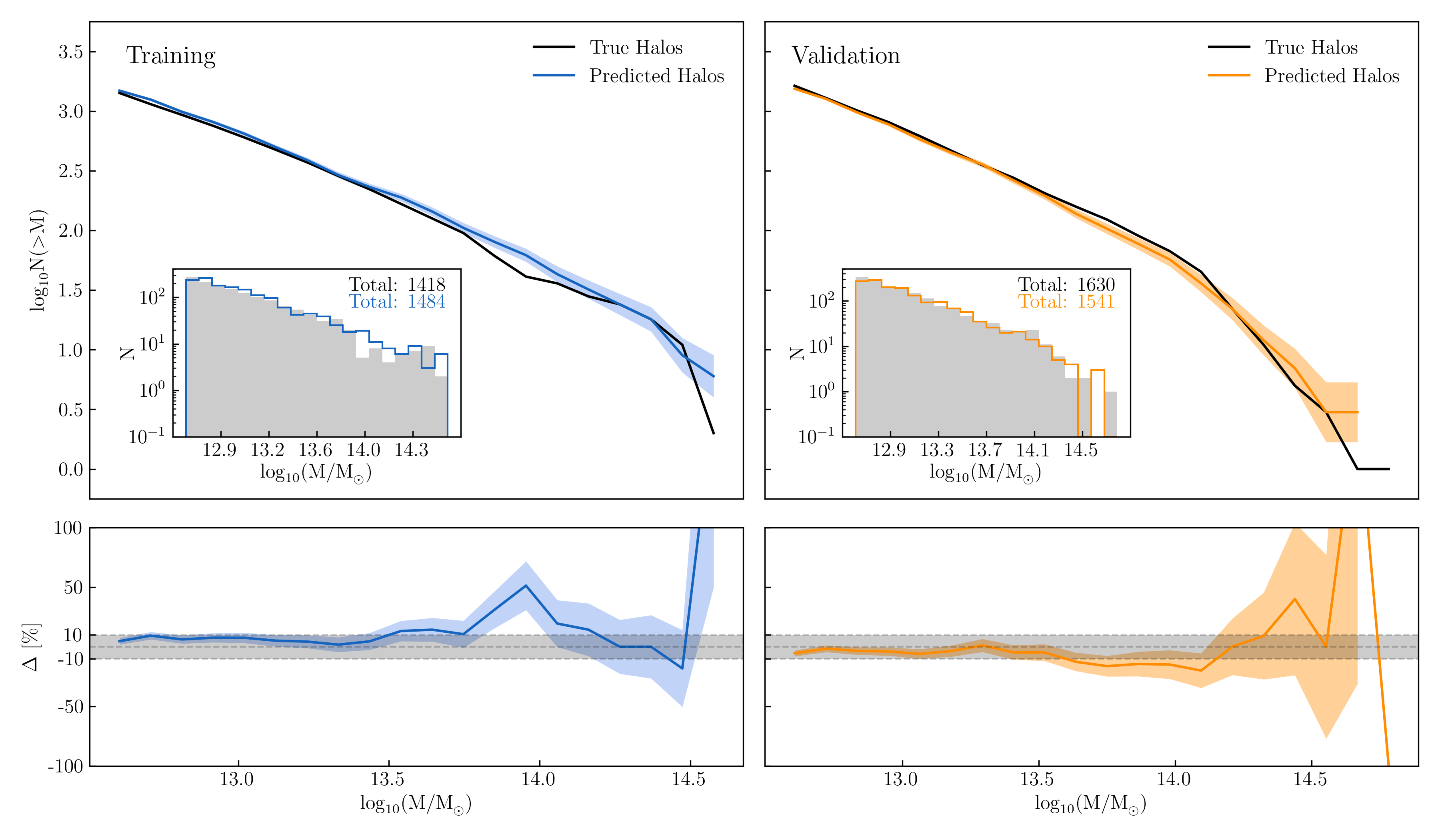}
 \caption{Reconstructed cumulative halo mass functions for the training (left) and validation (right) simulations at $z=0$, where N(>M) is the number of haloes with a mass greater than M. The corresponding ground truths are shown in black and the color shaded regions denote the Poisson uncertainty in the corresponding mass bin. Also shown as insets are the differential versions with the predicted and true halo number counts as well as the percentage deviation from the ground truth halo mass functions.}
 \label{fig:hmfT}
\end{figure*}

The collection of reconstructed protohalo sizes is then converted to the final mass catalogue by means of the background density at the ICs, i.e. $M=\bar{\rho}V$. This strategy works very well in the linear regime ($\delta \approx 0$). In fact, as the constructed grid matches the particle resolution of $512^{3}$, there is approximately one particle in each grid cell, such that final halo masses at $z=0$ equal the protohalo masses in very good agreement
. Thus, predicting the protohalo masses allows to reconstruct the halo mass function of dark matter haloes at $z=0$. In figure \ref{fig:hmfT} we show the true and predicted accumulated halo mass functions as well as the percentage deviation from the ground truth for the training and validation simulation. The reconstructed statistics match the ground truth catalogs for most mass bins within an uncertainty of 10\%. The deviation is largest for very massive protohalo regions of the order $\sim$10$^{14}\Msun$ and beyond. Given that the total numbers of these large scale objects in the entire simulation box is comparatively low, the variety of cases that the neural network is presented with during training is especially limited on these scales. We retrieve a total of 1484 protohalo regions for the training simulation and 1541 thereof in the validation case (5\% deviation in both cases). The neural network as well as the piece-wise linear correction fit generalize well to unseen data samples from the validation set (figure \ref{fig:triple_combined} and \ref{fig:fit}).

\section{Conclusions and future work}
\label{sec:conclusion}
We have presented a framework for the task of identifying protohalo regions on a one-by-one basis directly from the initial conditions of N-body simulations. We designed this challenging task as a hybrid approach consisting of pure Deep Learning (U-net) paired with an image reconstruction technique (watershed algorithm). We showed that by reformulating the underlying classification task as a distance-based regression problem, the comparatively small network with $\sim$14.5$\cdot10^{6}$ trainable parameters is in fact able to retrieve collapsing regions despite being trained on data samples from only one simulation box.\par
We argued that the native marker-based watershed algorithm is not sensitive enough for reconstructing the exact halo masses as it generally overpredicts the sizes of individual regions due to blindly flooding the 3D image. We implemented a computationally fast addition to complement the native algorithm with an adaptive post-processing element that is calibrated with the training simulation.  
We find that this hybrid model is able to reconstruct the halo mass function at $z=0$ within an uncertainty of $\sim$10\% for most mass bins. In addition to the statistical precision, the model also achieves good agreement when comparing the actual sizes and positions of protohalo regions on a one-by-one basis as seen in figure \ref{fig:triple_combined}. The hybrid model is designed to be robust against many complications that arise from fully non-linear dynamics in N-body simulations. The pixel-wise euclidean distance map as the target is completely free from any assumptions on the cluster morphology (e.g. spherical approximation), and bypasses the difficulties of training a multi-label classification task, where each protohalo region would be treated as a different class. Generally, the network achieves good precision on the regressed distance information and is able to isolate distinct protohalo regions.\par
However, a weakness of the current pipeline is that it occasionally fails to predict halo formation if multiple density patches ending up in a single halo originate from disconnected regions (as seen in case A in figure \ref{fig:triple_combined}). It is possible that these faulty predictions originate from cases where the density configuration alone might not yield enough information regarding how patches merge or split. For this problem, we propose to include the velocity field as an additional information carrier when training the neural network as it could provide missing information on the fragmentation of individual density patches. The current setup can easily be extended to predict more halo properties since this technique only depends on how the target fields are constructed numerically.
Future installments will investigate to what extent an improved hybrid model can predict merger scenarios based on information about the locations of overdensities and their relative velocities and gradients thereof. We expect that training deeper networks on larger datasets with higher resolution will decrease the systematic errors on the halo mass function. In our training strategy we made use of the normalized $L_{1}$ loss function, which measures deviations on a pixel-by-pixel level. Formulating a loss function that is more receptive and physically motivated regarding the underlying task of protohalo identification could offer a large opportunity for directing the learning process more towards the physics behind collapse dynamics \cite[e.g.][]{Karpatne2017}.\par
The problem of fully non-linear structure formation across a wide range of scales offers a challenging opportunity to test and fine-tune different machine learning approaches. From the results of this work, we conclude that Deep Learning models are capable of learning the relevant combinations from the input signals to identify collapsing regions in full N-body simulations. In this sense, the presented hybrid model offers a powerful base framework towards designing more precise algorithms with the eventual objective to build cosmic emulators for fast sampling of dark matter halo catalogues.

\section*{Acknowledgements}
RF acknowledges financial support from the Swiss National Science Foundation (grant no 157591).
\newpage



\bibliographystyle{mnras}
\bibliography{References}





\bsp	
\label{lastpage}
\end{document}